\documentclass[EN]{rcftex}

\title{Symmetry in the system conformed by two blocks connected by a string with variable tension} 
\vspace{1 cm}

\titleEN{Simetr\'Ia en dos bloques conectados por una cuerda con tensi\'on variable.}

\author{D. Benitez\toaff{a}, H. J. Herrera Su\'{a}rez\toaff{{b}\emailto} y J. H. Mu\~{n}oz\toaff{c}}

\affiliations{
 \aff  Instituto de educaci\'{o}n y pedagog\'{i}a, Universidad del Valle, Ciudad Universitaria Meléndez Calle 13 N\'umero 100-00, Cali, Colombia
\aff  Facultad de Ciencias Naturales y Matem\'{a}ticas, Universidad de Ibagu\'{e}, Carrera 22 Calle 67, barrio Ambal\'{a},Ibagu\'{e}, Colombia$^\emailto$. hernan.herrera@unibague.edu.co
\aff  Departamento de F\'{\i}sica, Universidad del Tolima, A. A. 546, Ibagu\'{e}, Colombia
}

\abstractES{In this paper we review the system made up by two blocks connected by a string over a smooth pulley with variable tension. One block lies on an horizontal surface and the another block is hanging vertically. We carry out a complete and systematic analysis for the tension of the string as function of the angle $\theta$ and the horizontal distance $x$, at static equilibrium. We find a simmetry-like that corresponds to two different configurations with the same tension and obtain the relationship that must satisfy two angles or two horizontal distances to obtain equal tension.}

\abstractEN{En este art\'iculo revisamos el sistema conformado por dos bloques conectados por una cuerda sobre una polea sin fricci\'on con tensi\'on variable. Un bloque se encuentra sobre una superficie horizontal y el otro est\'a suspendido verticalmente. Realizamos un an\'alisis completo y sistem\'atico para la tensi\'{o}n de la cuerda en funci\'{o}n del \'angulo $ \theta $ y la distancia horizontal $x$, en equilibrio est\'atico. Encontramos un tipo de simetr\'{i}a que corresponde a dos configuraciones diferentes con la misma tensi\'on y obtuvimos la relaci\'on que deben satisfacer dos \'angulos o dos distancias horizontales para obtener igual tensi\'on.  
}

\keyW{Newtonian mechanics (Mecánica newtoniana) 45.20.D; Tension measurement (medida de tensión) 07.10.Lw; Mechanical instrument equipment (equipo de instrumentos mecánicos) 07.10.-h.}

\begin{document}

\maketitle

\section{Introduction}

In this work we re-examine  the system of two blocks of masses $m$ and $M$ connected by a string over a smooth pulley, at static equilibrium. The string is extensionless, uniform and its  mass is negligible, and there is a coefficient of static friction between the mass $m$ and the horizontal surface. In Figure 1 we show the forces acting on this problem.

This system and similar versions are considered in fundamental physics textbooks \cite{Serway, Gonz, Ohanian, Hibbeler, Riley}, in some papers \cite{Lerman, Mak, Sutt, van, Leonard} and the website of A. Franco \cite{Franco}. However, a complete analysis about the tension $T$ in function of the angle $\theta$ or the horizontal distance $x$ has not been considered in the literature.

The mentioned problem is important because the tension $T$ and the normal force $N$ are not constant, unlike what happens in the following systems: the Atwood{'}s machine, a mass on  an horizontal surface and the another mass suspended, a mass on an inclined plane and the another mass hanged, and two masses on two inclined planes.

The purpose  of this work is to perform a complete analysis on the tension of the string, $T$, which is provided by the hanging mass $M$  when the system is at static equilibrium, as function of the angle $\theta$ and the horizontal distance $x$, using trigonometric functions and elementary mathematical tools. 

The paper is organized as follows. In section II we show the experiment arrangement built by us to model the system shown in Figure 1; in section III we present the theoretical analysis and the conclusions are given in section IV.

\begin{figure}
    \centering
    \includegraphics[width=0.4 \textwidth]{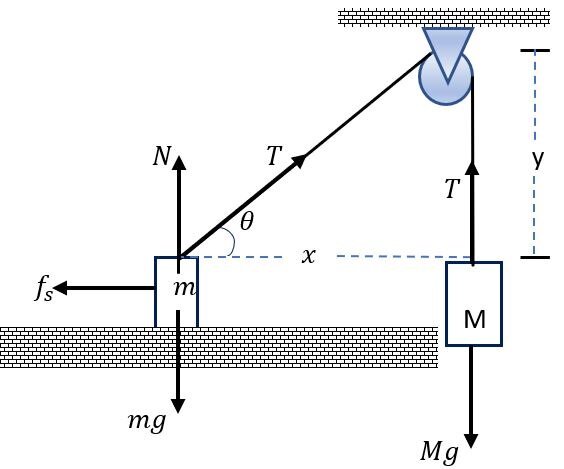}
    \caption{Two blocks tied to an extensionless rope. The mass of the string is negligible.}
\end{figure}

\section{The experiment}
Considering that only the reference \cite{van} performed the experimental arrangement and took experimental data, corresponding to the system showed in Figure 1, and assuming  that it is important because it articulates theory and experiment, we also carry out another arrangement. We show, in Figure 2, the experimental arrangement used for obtaining the tension $T$ in function of the angle $\theta$. We use the linear air track, reference $337501$, of the company Leybold \cite{Leybold}. We take $m=1.3745$ kg and $M= 0.4$ kg, and put a digital dynamometer \cite{amazon}  over the mass $m$ and the direction of the rope to measure the tension $T$. The mass of the dynamometer has been added to $m$. The angle $\theta$ was measured with the Angle Meter PRO+ free application of Play Store \cite{play}. 

We obtained that the static friction coefficient between the mass $m$ and the surface is $\mu_{s}=\frac{M}{m}=0.29$. We show in Figure 3 our experimental results for the tension $T$ in function of the angle $\theta$, considering several static equilibrium configurations. The estimated errors in the measurements of the angle $\theta $ and the
masses are 1$^{0}$ and 1 gr, respectively. The uncertainty of the tension $T$
is obtained in quadrature:%
\[
dT=\sqrt{\left( \frac{\partial T}{\partial \theta }\right) ^{2}\left(d\theta \right) ^{2}+\left( \frac{\partial T}{\partial M}\right) ^{2}\left(
dM\right) ^{2}+\left( \frac{\partial T}{\partial m}\right) ^{2}\left(dm\right) ^{2}}\], given approximately $\Delta T= 0.02 N$.
In the next section we present the theoretical analysis about this figure.  

\begin{figure}
    \centering
    \includegraphics[width=0.5 \textwidth]{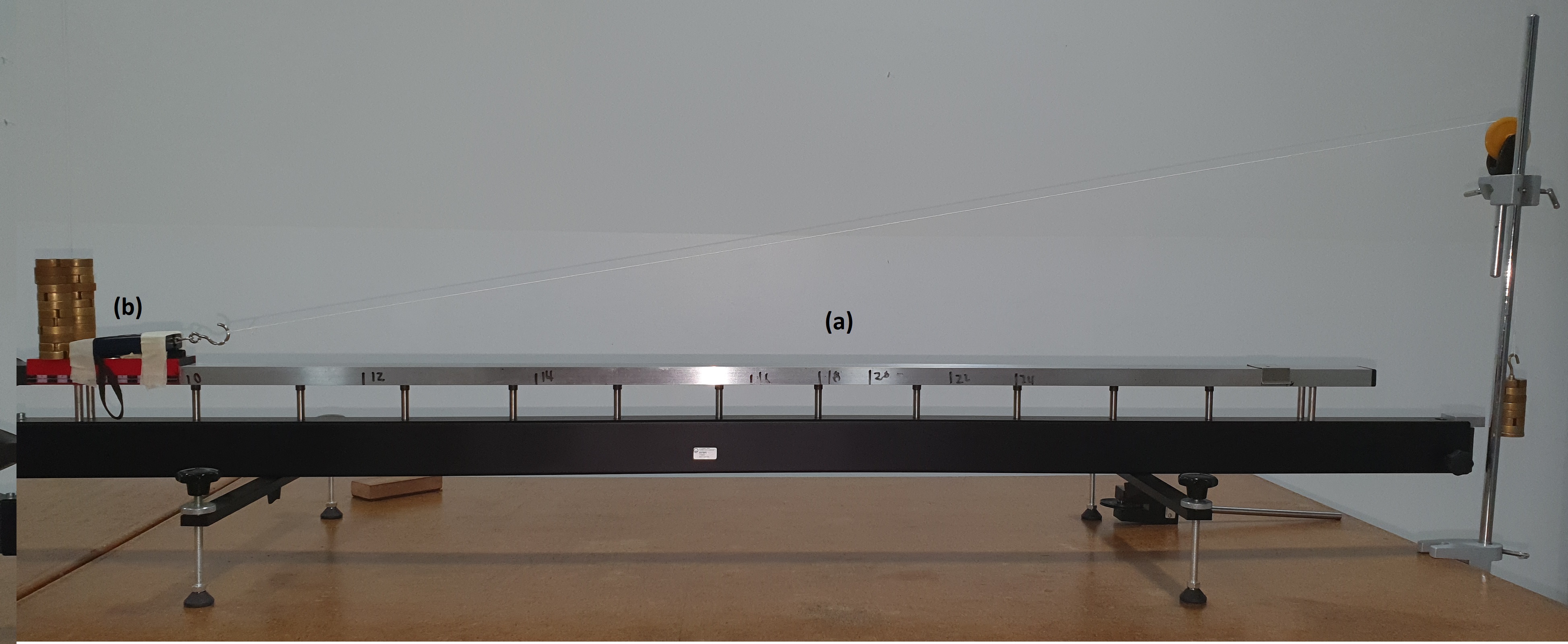}
    \caption{Experimental arrangement: (a) the linear air track; (b) the dynamometer}
\end{figure}

\begin{figure}
    \centering
    \includegraphics[width=0.5 \textwidth]{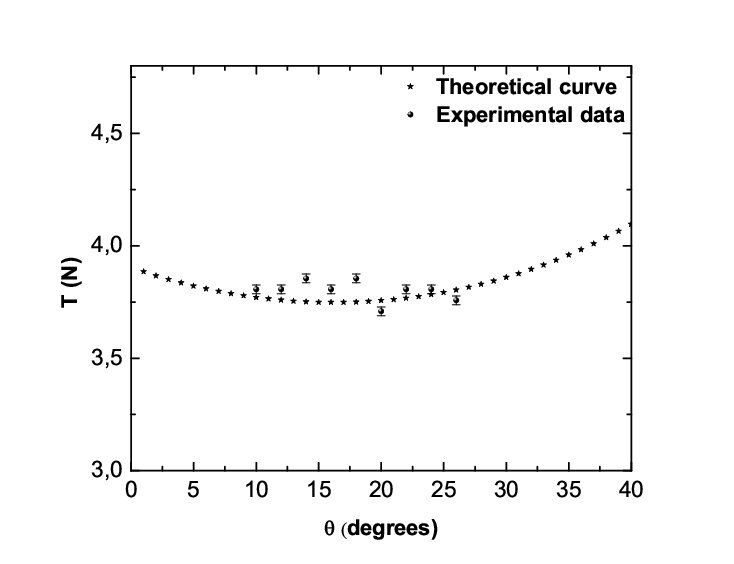}
    \caption{Experimental results for the tension $T$ in function of $\Theta$}
\end{figure}

\section{Theoretical analysis}

Applying the Newton{'}s second law to the mass $m$ and assuming that the system is at static equilibrium, the tension of the string,  $T$, in function 
the angle $\theta $ is given by \cite{Serway, Gonz, Lerman, Mak, Sutt, van, Franco}    
\begin{eqnarray}
T\left( \theta \right) =\frac{\mu _{s}mg}{\cos \theta +\mu _{s}\sin \theta },
\label{1}
\end{eqnarray}
\noindent where $\mu _{s}$ is the coefficient of static friction, $g$ is the gravity acceleration and  $\theta$ is in the range  $[0, \pi/2]$.The theoretical curve shown in Figure 3 was made with this equation. We can see that the experimental data agree, approximately, with the theoretical prediction. 

The minimum value of this tension is reached when

\begin{center}
$\theta_{min}=tan^{-1}(\mu_{s})$. 
\end{center}

\begin{figure}
  \centering
  \includegraphics[width=0.5\textwidth]{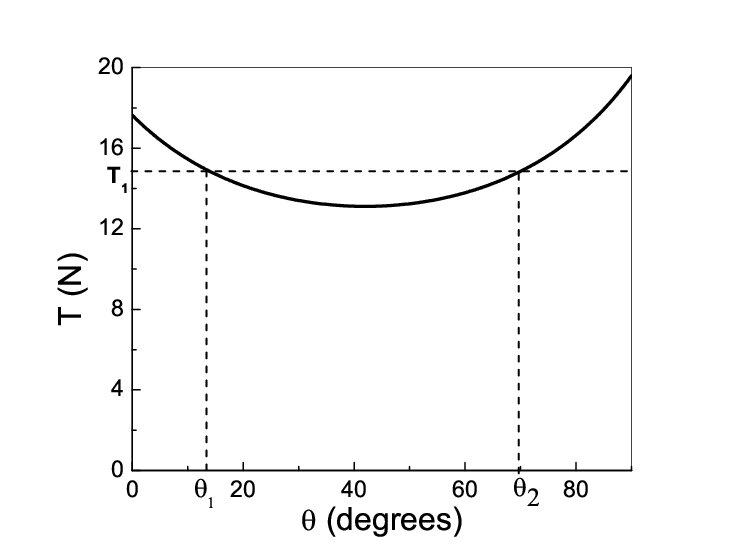}
  \caption{Tension $T$ in function of the angle $\theta $, with $m= 2$ kg, $y=0.3$ m and $\mu _{s}=0.9.$}
\end{figure}

This relation can be associated, in a pedagogical way, with the right triangle shown in Figure 5. By means of this figure it is easy to obtain
\begin{center}
$T_{min} = T(\theta_{min})=\frac{\mu_{s}mg}{\sqrt{1+\mu_{s}^{2}}}.$  
\end{center}

We display, in Figure 4, the tension $T(\theta)$ given by equation (1), with $m=2 kg$ and $\mu_{s}=0.9$. In this Figure it is possible to identify \textit{two extreme cases}: (1) When the string is parallel to the horizontal surface. In this case,  $\theta \rightarrow 0$, then 
$\lim\limits_{\theta \rightarrow 0}T\left( \theta \right) =\allowbreak \mu_{s}gm$;  (2)  When $\theta \rightarrow \pi/2$, then $\lim\limits_{\theta \rightarrow \frac{\pi }{2}}T\left( \theta \right) =mg$. In this situation the rope is perpendicular to the horizontal surface and  corresponds to the Atwood's machine at rest or at uniform motion.

\subsection{T as function of angle}

Let  us consider an arbitrary tension $T=T_{1}$ and draw a parallel line to the horizontal axis at height $T_{1}$ in Figure 4. Clearly, we see that there are two angles ($\theta_{1}$ and $\theta_{2}$) for which the same tension is obtained . It means that for each selected angle $\theta_{1}$ there is another angle $\theta_{2}$ that $T(\theta_{1})= T(\theta_{2})$, i.e., there are two identical configurations for the static equilibrium (see Figure 6), indicating certain type of symmetry in the system. One of the aims  of this  note is to find the restriction that these angles must follow.  From equation $\ref{1}$, we obtained 
\begin{eqnarray}
\frac{\mu _{s}mg}{\cos \theta _{1}+\mu _{s}\sin\theta_{1}}=\frac{\mu _{s}mg}{\cos \theta _{2}+\mu _{s}\sin \theta _{2}}  \label{3}.
\end{eqnarray}

Making a simple algebraic manipulation, we get

\begin{eqnarray}
\mu _{s}=\frac{\cos \theta _{1}-\cos \theta _{2}}{\left( \sin \theta
_{2}-\sin \theta _{1}\right) },  \label{5}
\end{eqnarray}

\begin{figure}
  \centering
  \includegraphics[width=0.45 \textwidth]{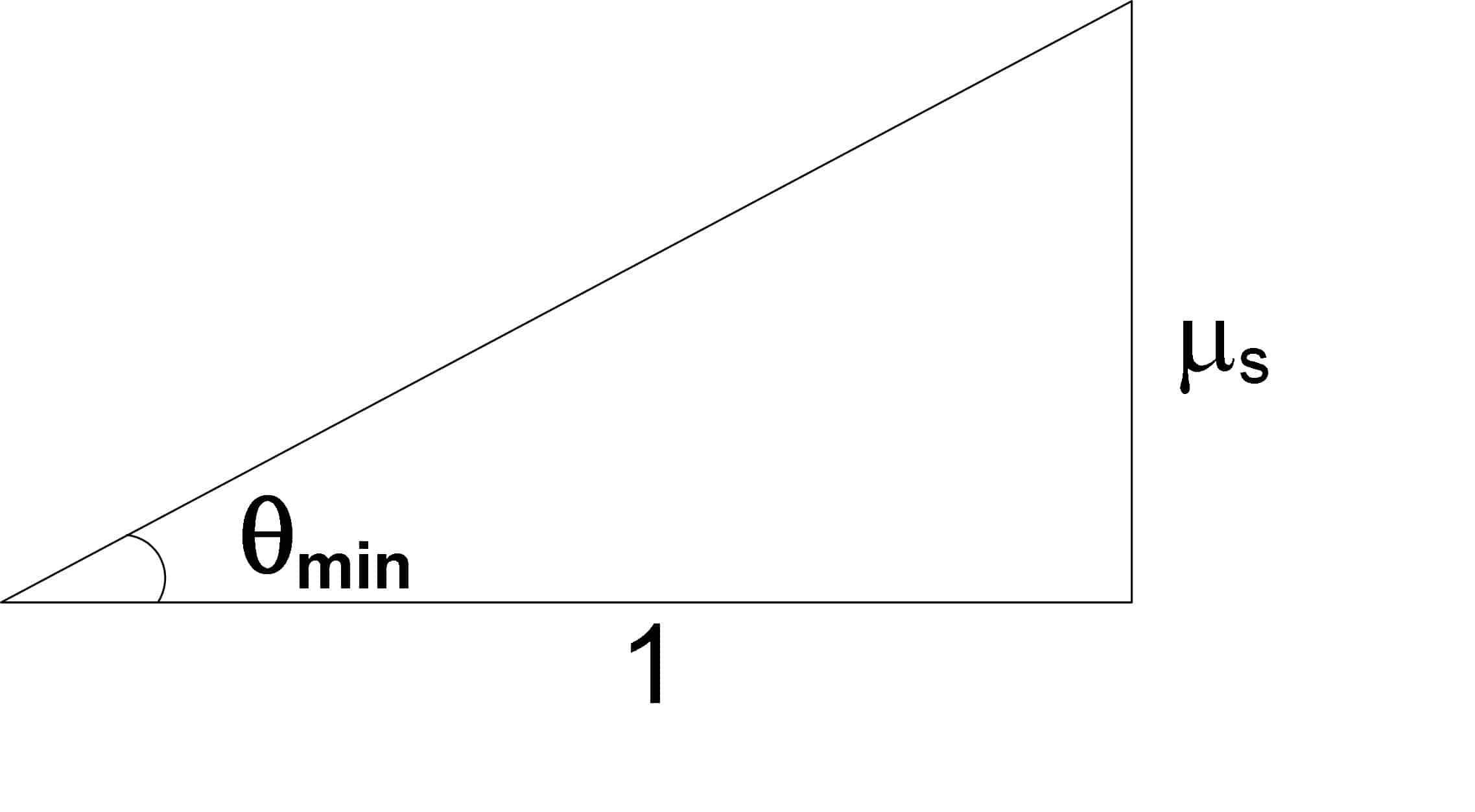}
  \caption{\ Right triangle associated to the relation $tan(\theta_{min})= \mu_s$}
\end{figure}

and using the well known trigonometric identities, 
\begin{eqnarray}
\cos A-\cos B &=&-2\sin \left( \frac{A+B}{2}\right) \sin \left( \frac{A-B}{2}%
\right) ,  \label{6} \\
\sin A-\sin B &=&2\cos \left( \frac{A+B}{2}\right) \sin \left( \frac{A-B}{2}%
\right) ,  \label{7}
\end{eqnarray}%

we can rewrite the equation $\ref{5}$ as,%

\begin{eqnarray}
\mu _{s} &=&\frac{-2\sin \left( \frac{\theta _{1}+\theta _{2}}{2}\right)
\sin \left( \frac{\theta _{1}-\theta _{2}}{2}\right) }{-2\cos \left( \frac{%
\theta _{1}+\theta _{2}}{2}\right) \sin \left( \frac{\theta _{1}-\theta _{2}%
}{2}\right) },  \nonumber \\
\mu _{s} &=&\tan \left( \frac{\theta _{1}+\theta _{2}}{2}\right) ,  \label{8}
\end{eqnarray}%
finally getting 
\begin{eqnarray}
\theta _{2}+\theta _{1}=2\tan^{-1} \left( \mu _{s}\right). 
\label{9.1}
\end{eqnarray}%
or equivalently 
\begin{eqnarray}
\theta _{2}+\theta _{1}=2\theta_{min}. 
\label{9}
\end{eqnarray}%
This equation establishes the relationship between the angles $\theta _{1}$ and $\theta _{2}$ that give the same tension. It provides us certain type of symmetry in this system.

\begin{figure}
  \centering
  \includegraphics[width=0.4 \textwidth]{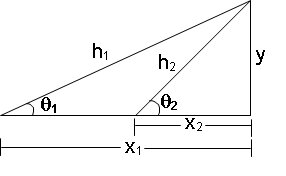}
  \caption{Configurations for the static equilibrium.}
\end{figure}
\vspace{1cm}

\subsection{T as function of x}

It is also possible analyze the tension $T$ in function of the horizontal distance $x$. For that, we use the following relations 
\begin{eqnarray}
\cos \theta &=&\frac{x}{\sqrt{x^{2}+y^{2}}},  \label{10} \\
\sin \theta &=&\frac{y}{\sqrt{x^{2}+y^{2}}},  \label{11}
\end{eqnarray}%
in the equation (\ref{1}) and obtained 
\begin{eqnarray}
T\left( x\right) =\allowbreak \frac{\mu _{s}gm\sqrt{x^{2}+y^{2}}}{x+y\mu _{s}%
}.
\label{Tx}
\end{eqnarray}
A graph of this equation is plotted in Figure 7. 

Again, if we draw a parallel line to the horizontal axis at height $T_{2}$  we see that there are two different positions $x_1$ and $x_2$ that yield $T(x_1) = T(x_2)$ i.e., there are two similar configurations of static equilibrium (see Figure 7). The minimum value for $T$ is obtained when
\begin{center}
$x_{min}=\frac{y}{\mu_s}$. 
\end{center}
In this situation is obtained  
\begin{center}
$T_{min} = T(x_{min}) = \frac{\mu_{s}mg}{\sqrt{1+\mu_{s}^{2}}}.$ 
\end{center}

Thus, $T(x_{min})=T(\theta_{min})$.

On the other hand, we can find the \textit{two extreme cases} considered in the previous subsection: (i) We obtained  that $\lim\limits_{x\rightarrow 0}T\left( x\right) =\allowbreak mg$. This result  is equivalent to $\lim\limits_{\theta \rightarrow \frac{\pi }{2}}T\left( \theta \right)$; (ii) Furthermore $\lim\limits_{x\rightarrow \infty }T\left( x\right)=\allowbreak \mu _{s}mg$. This situation agrees with $\lim\limits_{\theta \rightarrow 0}T\left( \theta \right) $.

Now we are going to find the condition that must satisfy the horizontal positions $x_1$ and $x_2$ that yield $T(x_1)=T(x_2)$. We obtain this relation in three different forms:
\begin{itemize}

    \item  \textit{First:} From equation (\ref{Tx}) is obtained

\begin{eqnarray}
\left( \allowbreak T^{2}-g^{2}m^{2}\mu _{s}^{2}\right) x^{2}+2T^{2}y\mu
_{s}x+\left( T^{2}-\allowbreak g^{2}m^{2}\right) y^{2}\mu _{s}^{2}=0,
\label{15}
\end{eqnarray}%
Resolving this quadratic equation in $x$, we get

\begin{equation}
x_{1} =\allowbreak \frac{-T^{2}y\mu _{s}+gmy\mu _{s}\sqrt{T^{2}\mu
_{s}^{2}+T^{2}-g^{2}m^{2}\mu _{s}^{2}}}{T^{2}-g^{2}m^{2}\mu _{s}^{2}}
\label{16} \\
\end{equation}%

\begin{equation}
x_{2} =\allowbreak \frac{-T^{2}y\mu _{s}-gmy\mu _{s}\sqrt{T^{2}\mu
_{s}^{2}+T^{2}-g^{2}m^{2}\mu _{s}^{2}}}{T^{2}-g^{2}m^{2}\mu _{s}^{2}}.
\label{17}
\end{equation}%

\begin{figure}
  \centering
  \includegraphics[width=0.45 \textwidth]{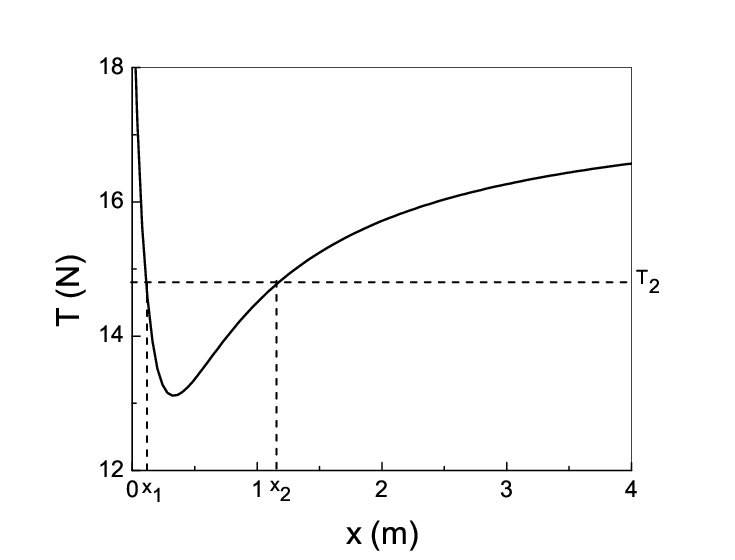}
  \caption{Tension $T$ as function of $x$, with $m= 2$ kg, $y=0.3$ m and $\mu _{s}=0.9.$ }
\end{figure}

The sum of these roots is
\begin{eqnarray}
x_{1}+x_{2}=\frac{2T^{2}y\mu _{s}}{g^{2}m^{2}\mu _{s}^{2}-T^{2}}.  \label{Hernan}
\end{eqnarray}
From this equation, $x_{2}$ can be obtained knowing $x_{1}$, the tension $T$, the coefficient of static friction $\mu_{s}$ and the vertical distance $y$.

\item \textit{Second:} From Figure 6 we can build Figure 8 and after applying the Law of Sines to obtain:
\begin{eqnarray}
\frac{\sin \theta _{1}}{\sqrt{x_{2}^{2}+y^{2}}}=\frac{\sin \theta _{2}}{%
\sqrt{x_{1}^{2}+y^{2}}},
\end{eqnarray}
or
\begin{eqnarray}
x_{1}^{2}\sin ^{2}\theta _{1}-x_{2}^{2}\sin ^{2}\theta _{2}=y^{2}\left( \sin
^{2}\theta _{2}-\sin ^{2}\theta _{1}\right).
\end{eqnarray}

\begin{figure}
  \centering
  \includegraphics[width=0.45 \textwidth]{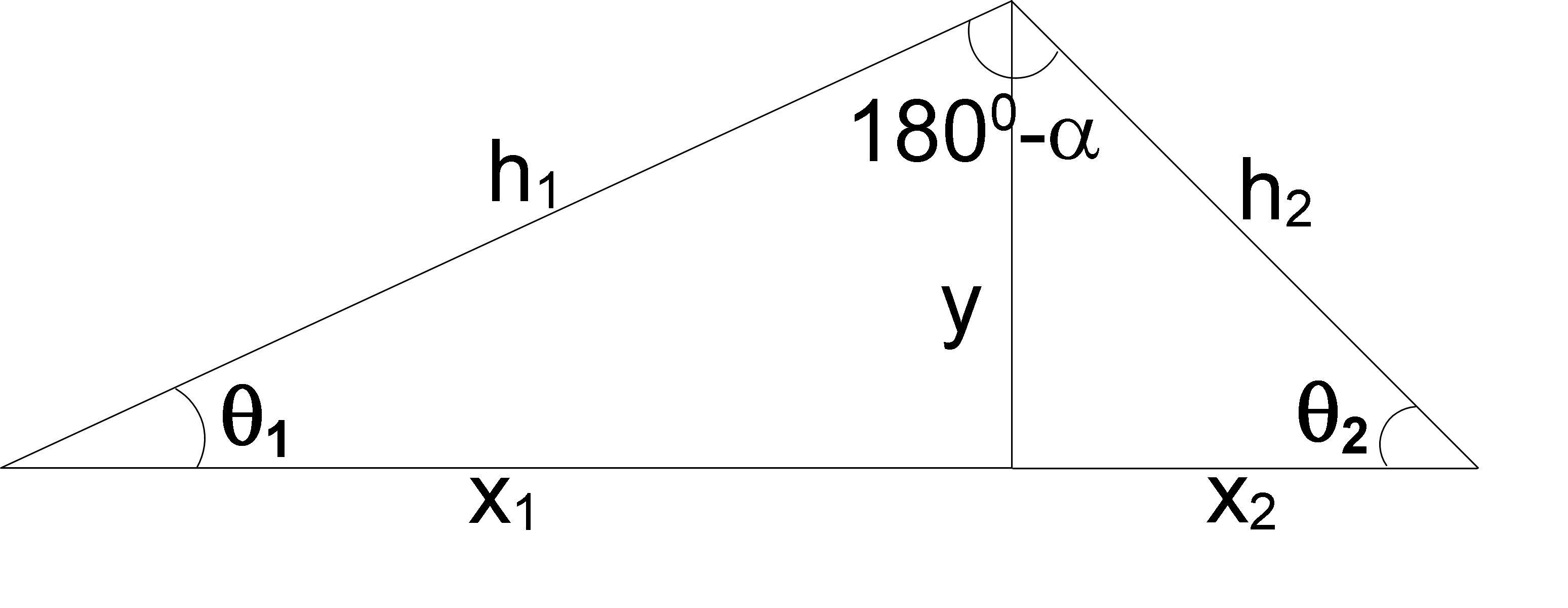}
  \caption{Reinterpretation of Figure 6.}
\end{figure}

Using the trigonometric identity $\sin ^{2}\left( A\right) -\sin ^{2}\left( B\right) =\sin \left( A+B\right)
\sin \left( A-B\right)$, it is obtained
\begin{eqnarray}
x_{1}^{2}\sin ^{2}\theta _{1}-x_{2}^{2}\sin ^{2}\theta _{2}=y^{2} \sin
\left( \theta _{1}+\theta _{2}\right) \sin \left( \theta _{1}-\theta
_{2}\right) 
\end{eqnarray}
or
\begin{eqnarray}
x_{1}^{2}\sin ^{2}\theta _{1}-x_{2}^{2}\sin ^{2}\theta _{2}=y^{2}\sin \left(
2\theta _{\min }\right) \sin \left( \theta _{1}-\theta _{2}\right). 
\label{Herman}
\end{eqnarray}
From this equation, we can obtained $x_{2}$ knowing $x_{1}$ and the angles $\theta_{min}$ and $\theta_{1}$.

\item \textit{Third:} If $x_{2}=kx_{1}$,  $k\in \Re ^{+}$: using $\tan \theta _{1}=\frac{y}{x_{1}}$, $\tan \theta _{2}=\frac{y}{x_{2}}$ in  the trigonometric identity   
\begin{eqnarray}
\tan \left( \theta _{1}+\theta _{2}\right) =\frac{\tan \theta _{1}+\tan
\theta _{2}}{1-\tan \theta _{1}\tan \theta _{2}},
\end{eqnarray}

the following quadratic equation in $x_1$ is obtained:
\begin{equation*}
k\tan \left( 2\theta_{min} \right) x_{1}^{2}-y\left( k+1\right) x_{1}-y^{2}\tan
\left( 2\theta_{min} \right) = 0,
\end{equation*}
which has  the physical solution 
\begin{equation}
x_{1}=y\left( \frac{\left( k+1\right) +\sqrt{\left( k+1\right) ^{2}+4k\tan
^{2}\left( 2\theta_{min} \right) }}{2k\tan \left( 2\theta_{min} \right) }\right).   \label{David}
\end{equation}

\end{itemize}

From this equation, it can be obtained $x_{1}$ for a given $k$ and knowing the vertical distance $y$ and the $\theta_{min}$ angle.

\section{Concluding remarks}
We examined the static equilibrium problem of a block of mass $m$ on a plane being pulled at an angle $\theta$ with the horizontal by a tension due to a suspended mass $M$ from a pulley. We performed a complete analysis to the expression for the tension of the string, $T$, when the system is at static equilibrium. We found two different configurations with equal tension showing certain kind of symmetry, and showed that there are two extreme cases that can be related with the Atwood's machine and the system conformed by one mass on an horizontal surface connected with another mass  suspended vertically from a pulley.

First, we analyzed the tension $T$ in function of the angle $\theta$. We found (except for $ \theta_{min}$= $\tan^{-1} \left( \mu _{s}\right)$, where $\mu _{s}$ is the coefficient of static friction) that for each angle $\theta_1$ there is another configuration of static equilibrium given by the angle $\theta_2$ so that $T(\theta_1)=T(\theta_2)$ with the condition  $\theta_1+\theta_2=2tan^{-1}\mu_s=2\theta_{min}$. Second, we analyzed, in a similar way, the tension $T$  in function of the horizontal distance $x$. Again, we obtained that there are two horizontal distances  $x_1$ and $x_2$ that give a configuration of the static equilibrium such that $T(x_1)=T(x_2)$. The condition that must satisfy $x_1$ and $x_2$ is given, in three different forms, by means of the equations (\ref{Hernan}), (\ref{Herman}) and (\ref{David}). According to our knowledge, these results have not yet been published. 

Additionally, we did a nice laboratory exercise in order to check experimentally the 
 equation (1). The analysis carried out in this work is an important didactic tool that allows to articulate physics and mathematics, and theory with experiment, contributing to improve the learning of physics.

In this work we used simple and elementary mathematical tools as trigonometric identities and minima of a function, the free software Geogebra and the computer program Origin. Our results can be incorporated as additional questions to this problem in fundamental physics textbooks and introductory-level physics courses, and may help teachers to produce meaningful learning when teaching required, at the same time, the application of both trigonometric and statics.

 \textbf{Acknowledgments.}
 The authors acknowledgment to Juan Carlos Otavo for his collaboration in the experimental arrangement.



\end{document}